\documentclass[aps,prb,twocolumn,groupedaddress,showpacs]{revtex4}

\usepackage{graphicx}
\usepackage{dcolumn}
\usepackage{bm}

\begin{document}


\title{Landau model for the elastic properties of the ferroelastic crystal
Rb$_4$LiH$_3$(SO$_4$)$_4$.}


\author{G. Quirion, W. Wu\dag, J. Rideout}
\affiliation{Department of Physics and Physical Oceanography,
Memorial University, St. John's, Newfoundland, Canada, A1B 3X7
\\
\dag\ Department of Physics, University of Toronto, Toronto,
Ontario, Canada}
\author{B. Mr\`oz}
\affiliation{Institute of Physics, Adam Mickiewicz University,
Poznan, Poland}


\date{\today}

\begin{abstract}
We report a detailed investigation of the elastic properties of
Rb$_4$LiH$_3$(SO$_4$)$_4$ measured as a function of temperature and
pressure. Rb$_4$LiH$_3$(SO$_4$)$_4$ is known to show a $4
\rightarrow 2$ ferroelastic phase transition at T$_c = 134$~K.  In
order to clarify the nature of the order parameter associated with
that structural transition, we compare our finding to two distinct
phenomelogical Landau models.  The coupling parameters of both
models are all determined using the temperature dependence of the
strain tensor, as well as the pressure dependence $d T_c / d P =
19.1 \pm 0.2~K/kbar$, prior to the calculation of the elastic
constants. Our comparison indicates that the ferroelastic transition
in Rb$_4$LiH$_3$(SO$_4$)$_4$ is fully consistent (within a few
percent) with the predictions of a pseudo-proper model, showing at
the same time that the primary mechanism leading to the phase
transition is not driven by strains. Our analysis also refutes the
idea that Rb$_4$LiH$_3$(SO$_4$)$_4$ is the first ferroelastic
compound showing incomplete softening, of the effective soft
acoustic mode $C_{soft}$, at T$_c$.

\end{abstract}


\maketitle

\section{Introduction}
Over the years, there has been a series of controversies regarding
the properties of the ferroelastic compound
Rb$_4$LiH$_3$(SO$_4$)$_4$. Initially designated as
LiRb$_5$(SO$_4$)$_2$1.5H$_2$SO$_4$ \cite{Mroz88, Mroz89, Wolejko88a,
Wolejko88b}, subsequent chemical analysis revealed that
Rb$_4$LiH$_3$(SO$_4$)$_4$ is the proper description. It was also
suggested that Rb$_4$LiH$_3$(SO$_4$)$_4$ belongs to the $4mm$ point
group in the paraelastic phase at ambient temperature. Consequently,
a $4mm \rightarrow 2mm$ symmetry change, associated with a
ferroelastic phase transition at 134~K, was
assumed\cite{Piskunowicz89, Minge88}. Additional
X-ray\cite{Pietraszko88} and neutron diffraction\cite{Mroz97}
measurements finally established that the proper group-subgroup
symmetry change is actually $4 \rightarrow 2$. Based on this new
phase sequence, the elastic properties of Rb$_4$LiH$_3$(SO$_4$)$_4$
have been revisited using Brillouin scattering\cite{Mroz91} and
ultrasonic\cite{Breczewski90} measurements. Both experimental
approaches now confirmed that the effective elastic constant
$(C_{11}-C_{12})/2$ softens as the temperature is reduced to $T_c$,
which is perfectly compatible with a $4 \rightarrow 2$ phase
transition. Nonetheless, in order to explain their respective
findings, each group proposed models which are fundamentally
different. On the one hand, the ultrasonic
measurements\cite{Breczewski90}, which show a sudden drop at $T_c$
in the velocity of longitudinal modes propagating along [100] and
[001], are interpreted within the framework of a pseudo-proper
ferroelastic model. Consequently, the authors assume that the order
parameter is not one of the strains or a combination of strain
components. However, in that case the order parameter must at least
exhibit the same symmetry as the strain $(e_1 - e_2)$ or
$e_6$\cite{Toledano83}. On the other hand, Mr\'oz et
al.\cite{Mroz91} claim that Rb$_4$LiH$_3$(SO$_4$)$_4$ shows
incomplete softening at $T_c$ and proposed a model which they claim
is consistent with that observation. In their paper,
Rb$_4$LiH$_3$(SO$_4$)$_4$ is described as a proper ferroelastic
compound where the strain combination $e_{soft} = \alpha_1~ (e_1 -
e_2) + \alpha_6~ e_6$ acts as the order parameter. Incomplete
softening in ferroelastic crystals is not a common phenomena, but it
has previously been observed in BiVO$_4$\cite{Pinczuk77}. In that
case, the softening is mediated by the coupling between optical and
acoustical modes. Considering that the model proposed by Mr\'oz et
al.\cite{Mroz91} includes no such coupling term, we believe that
their conclusion regarding the soft mode Rb$_4$LiH$_3$(SO$_4$)$_4$
requires further investigation. Thus, in order to clarify the nature
of the ferroelastic transition, as well as, to validate or refute
the possibility of incomplete softening in
Rb$_4$LiH$_3$(SO$_4$)$_4$, we present a detailed investigation of
the elastic properties of Rb$_4$LiH$_3$(SO$_4$)$_4$ as a function of
temperature and pressure. We also compare the predictions of a
proper and a pseudo-proper ferroelastic models in order to ascertain
the character of the ferroelastic transition in
Rb$_4$LiH$_3$(SO$_4$)$_4$.

\section{Experiment}
The Rb$_4$LiH$_3$(SO$_4$)$_4$ crystal was grown by the Crystal
Physics Laboratory of the Institute of Physics at Mickiewicz
University, Poland\cite{Mroz91}. For the ultrasonic investigation,
several samples in the form of cubes of about $3 \times 3 \times
3~mm^3$ were used to measure the sound velocity along different
crystallographic directions. Longitudinal and transversal waves were
generated and detected using 30~MHz lithium niobate transducers
mounted on one face of the crystal. The relative variations in the
sound velocity ($\Delta V / V$) were obtained using a
high-resolution pulsed acoustic interferometer as a function of
temperature and pressure. For measurements realized at high
pressure, the transducer-sample assemblage was inserted in a Cu-Be
pressure cell filled with a 3-Methyl-1-butanol fluid acting as the
pressure-transmitting medium. A small wire of Lead mounted close to
the sample was used during the experiments to determine the actual
pressure at different temperatures.

\section{Experimental Results}

 \begin{figure}[tb]
 \includegraphics{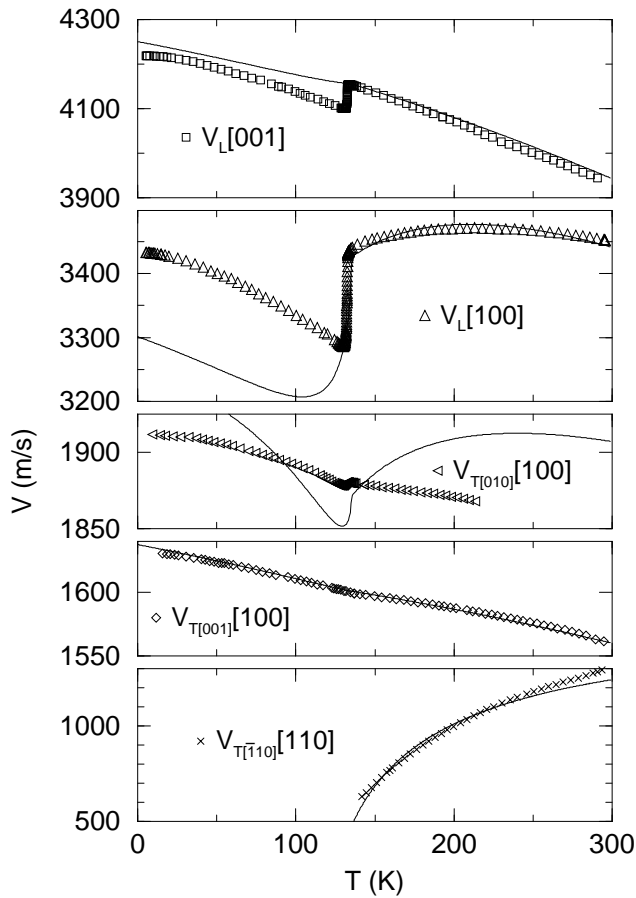}
     \caption{Temperature dependence of the sound velocity of longitudinal (L)
     and transversal (T) modes measured along the principal crystallographic directions
     of Rb$_4$LiH$_3$(SO$_4$)$_4$.
     The experimental results are represented by open symbols while the
     continuous lines correspond to predictions derived from the
     pseudo-proper ferroelastic model presented in the Appendix.
     \label{vvst} }
\end{figure}

\begin{figure}[tb]
 \includegraphics{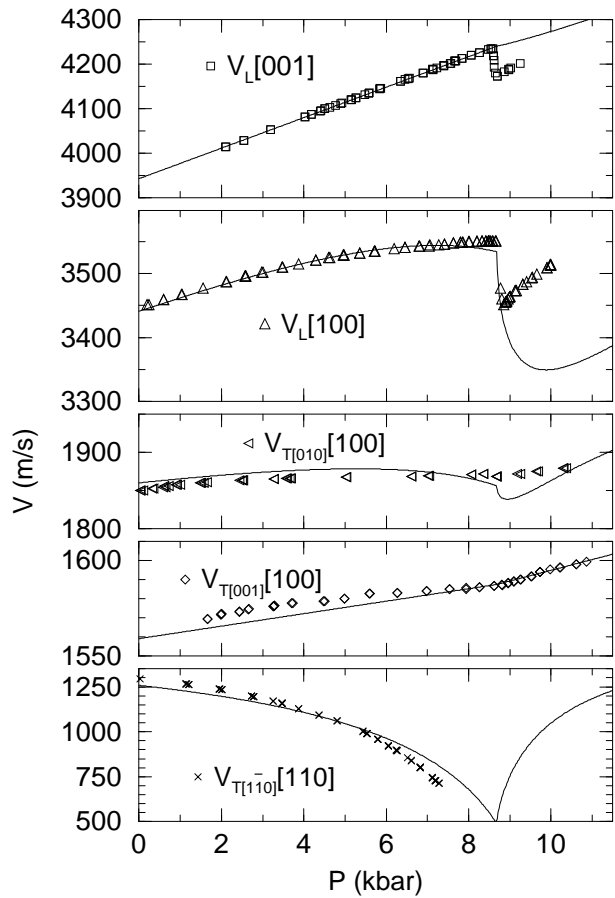}
     \caption{Pressure dependence of the sound velocity of longitudinal (L) and
     transversal (T) modes measured at room temperature along the principal
     crystallographic directions of Rb$_4$LiH$_3$(SO$_4$)$_4$.
     The experimental results are represented by open symbols while the
     continuous lines correspond to predictions derived from the
     pseudo-proper ferroelastic model presented in the Appendix.
     \label{vvsp} }
 \end{figure}

\begin{figure}[tb]
 \includegraphics{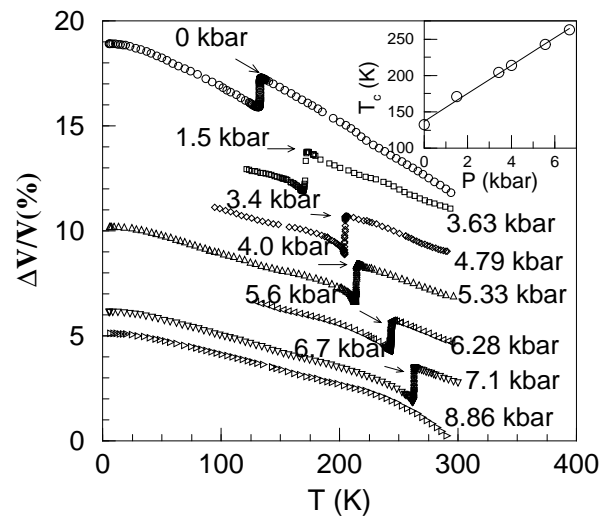}
     \caption{Temperature dependence of the sound velocity of longitudinal waves propagating
     along the z-axis ($V_L[001]$) measured at different pressures.
     The inset shows the pressure dependence of the ferroelastic critical
     temperature $T_c$ for Rb$_4$LiH$_3$(SO$_4$)$_4$. \label{C33atP} }
 \end{figure}

In this paper, the labels L and T are used to identify longitudinal and transverse modes,
respectively. In addition, the Miller index adjacent to the label T gives the
polarization of the mode while the other index corresponds to the direction of
propagation. For example, $V_{T[001]}[100]$ stands for the velocity of transverse wave
propagating along the crystallographic direction [100] with its polarization along [001].
Using that notation, we present in Fig.~\ref{vvst} and Fig.~\ref{vvsp} our high
resolution sound velocity results obtained for Rb$_4$LiH$_3$(SO$_4$)$_4$ as a function of
temperature and pressure using modes propagating along the principal axes. In these
figures, the experimental results are represented by open symbols while the continuous
lines correspond to predictions derived from the pseudo-proper ferroelastic model
presented in the Appendix~\ref{pseudo}. Let first point out that our results are very
consistent with previous studies\cite{Breczewski90, Mroz91} realized as a function of
temperature. As for the Brillouin scattering measurements\cite{Mroz91}, the results
presented in Fig.~\ref{vvst} indicate that the largest softening is observed for
transverse modes propagating along $[110]$ with a $[\overline{1}10]$ polarization
($V_{T[\overline{1}10]}[110]$). In addition, our high resolution measurements show how
the other modes change in the vicinity of the ferroelastic transition ($T_c = 132.8~K$).
In particularly, we note that the velocity of longitudinal modes presented in
Fig.~\ref{vvst} drop by a few percent at $T_c$.  This observation is also consistent with
previous ultrasonic measurements\cite{Breczewski90}. Our investigation is complemented
with the first measurements realized as a function of pressure which we presented in
Fig.~\ref{vvsp}. At ambient temperature, the data indicate a phase transition at a
pressure of $P_c = 8.6 \pm 0.2~kbar$. Considering that the observed anomalies on the
elastic modes at $T_c$ and $P_c$ are very similar, we can assume that the transition
observed at $P_c = 8.6 \pm 0.2~kbar$ corresponds to a $4 \rightarrow 2$ ferroelastic
structural transformation. To determine the pressure dependence of the critical
temperature, we have carried out a series of sound velocity measurements as a function of
temperature at different pressures. The results presented in Fig.~\ref{C33atP}
have been obtained using longitudinal waves propagating along the
z-axis. From these measurements, we find that the ferroelastic
transition temperature increases at a rate of $d T_c / d P = 19.1
\pm 0.2~K/kbar$ with pressure (see inset of Fig.~\ref{C33atP}).

\section{Data Analysis} \label{analysis}

\begin{figure}[tb]
 \includegraphics{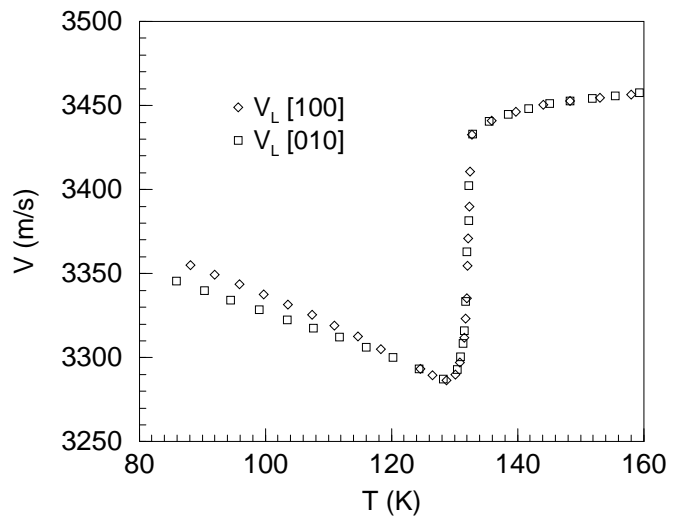}
     \caption{Comparison of the temperature dependence of the velocity of longitudinal
     waves propagating along two orthogonal directions in the ab plane ([100]
     and [010] directions).\label{V100vsV010} }
\end{figure}

The results presented in Fig.~\ref{C33atP} show that the temperature
dependence just below the critical temperature changes significantly
between ambient pressure and a moderate pressure of $P = 1.5$~kbar.
We believe that this might be an indication that the observed
temperature dependence in the ferroelastic state is not purely
intrinsic. One of the characteristic of ferroelastic compounds is
that structural domains appear in the ferroelastic phase. Moreover
for soft ferroelastic materials, it is possible to switch the
orientation of these domains by applying an uniaxial stress. In the
case of Rb$_4$LiH$_3$(SO$_4$)$_4$, the domains have been
observed\cite{Mroz97} and consist of two mutually perpendicular
walls. Considering that the pressure is not perfectly isotropic in
pressure cell, the unusual behavior of the velocity close to $T_c$
might be associated with a modification of the domain structure with
increasing pressure. Consequently, before attempting to analyze any
results in the ferroelastic phase, it is crucial to determine what
might be the influences of these domains on the measured quantities.
Thus, in order to ascertain if our measurements represent the
intrinsic elastic properties of Rb$_4$LiH$_3$(SO$_4$)$_4$, we
compare in Fig.~\ref{V100vsV010} the temperature dependence of the
velocity of longitudinal waves propagating along two orthogonal
directions in the ab plane ([100] and [010] directions).
As the $x$ and $y$ directions are no longer equivalent in a
monoclinic phase, one would expect a significant difference between
these two modes. The fact that both sets of data are practically
identical below $T_c$ indicates that the size of the domains is
smaller then our acoustic wavelength ($\lambda \approx 100~\mu m$)
at 30~MHz.  Thus, based on this comparison, it is clear that our
results obtained in the ferroelastic phase, as well as those
obtained by Breczewski et al. \cite{Breczewski90}, do not reflect
the intrinsic elastic properties of Rb$_4$LiH$_3$(SO$_4$)$_4$. For
that reason, the Landau models presented in the Appendix are tested
using results obtained principally in the paraelastic phase
(tetragonal phase).

So far, the elastic properties of Rb$_4$LiH$_3$(SO$_4$)$_4$ have
been qualitatively analyzed by different groups\cite{Mroz91,
Breczewski90} using two different Landau-type models. In one case,
they consider that the order parameter corresponds to the strains
$e_1 - e_2$ and $e_6$ (proper ferroelastic) while in the other
scenario the order parameter is unknown but has the same symmetry as
the strains mentioned previously (pseudo-ferroelastic). In this
paper, we present (see Appendix) a comprehensive derivation of a
proper and pseudo-proper models for a $4 \rightarrow 2$ ferroelastic
phase transition.  Both of these models are derived using a minimum
number of coupling factors. Our principal goal is to see if one
could determine the true nature of the order parameter based on a
detailed analysis of the elastic properties of
Rb$_4$LiH$_3$(SO$_4$)$_4$.

According to both models derived in the Appendix (see Eq.~\ref{CiiT}
and Eq.~\ref{CiiTproper}), the elastic constants $C_{11}$, $C_{16}$,
and $C_{66}$ are expected to soften while $C_{12}$ becomes stiffer
as $T_c$ is approached from above (or approach $P_c$ from below).
The main qualitative difference between these models is that the
softening and stiffening are expected to be linear for a proper
ferroelastic transition while it could be non-linear in the other
case. Using the relations given in reference [10] and our high
resolution velocity measurements, it is easy to determine the
temperature and pressure dependence of the principal elastic
constants. Considering that the value of $C_{16}$ is an order of
magnitude smaller than $C_{11}$\cite{Mroz91, Breczewski90}, we can
consider, as a reasonable approximation, that the temperature and
pressure dependence of $C_{11}$ is captured by that of $V_L[100]$.
Thus, our high resolution measurements indicate that $C_{11}$ indeed
softens, however, its temperature and pressure dependence are
significantly non-linear.
\begin{figure}[b]
 \includegraphics{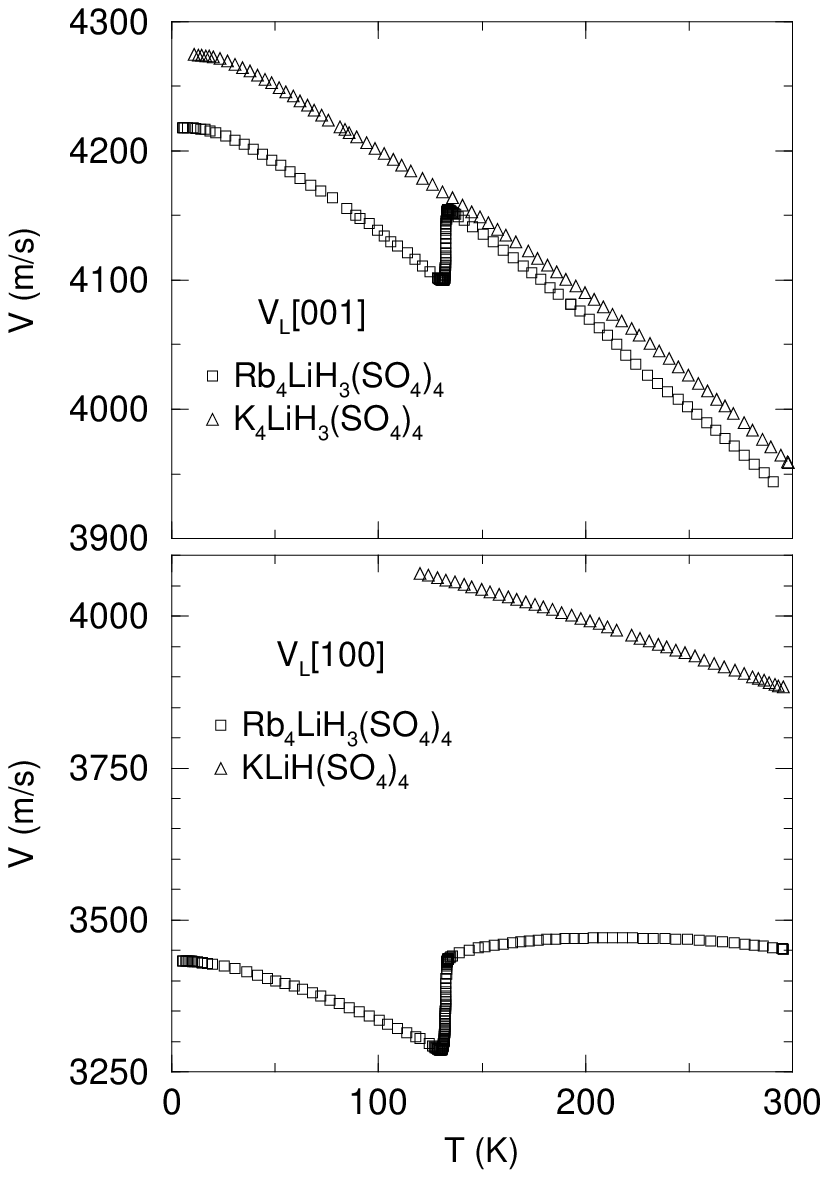}
     \caption{Comparison of the elastic properties of $Rb_4LiH_3(SO_4)_4$ and
     the isomorphic compound $K_4LiH_3(SO_4)_4$ using the temperature dependence
     of $V_L[100]$ and $V_L[001]$.\label{VRb4andK4} }
\end{figure}
This first constatation is reinforced by a comparison of the velocity measurements
performed on $Rb_4LiH_3(SO_4)_4$ and the isomorphic compound $K_4LiH_3(SO_4)_4$. As
$K_4LiH_3(SO_4)_4$ shows no ferroelastic transition\cite{Mroz91}, it can be used to
accurately determined the anharmonic temperature dependence of the elastic constants in
the absence of softening due to the transition. As an example, we compare in
Fig.~\ref{VRb4andK4} the results obtained for longitudinal modes propagating along the
$z$ and $x$ directions. Along the $z$-direction, we note that both compounds have the
same absolute velocity at room temperature and exhibit the same temperature dependence.
This is consistent with both models as no influence of the soft mode is expected on
$C_{33}$ in the para-elastic phase. Along the $x$-direction, the comparison of $V_L[100]$
for these two compounds indicates that the contribution associated with the soft mode is
indeed non-linear and thus more compatible with the pseudo-ferroelastic description.

\begin{figure}[tb]
 \includegraphics{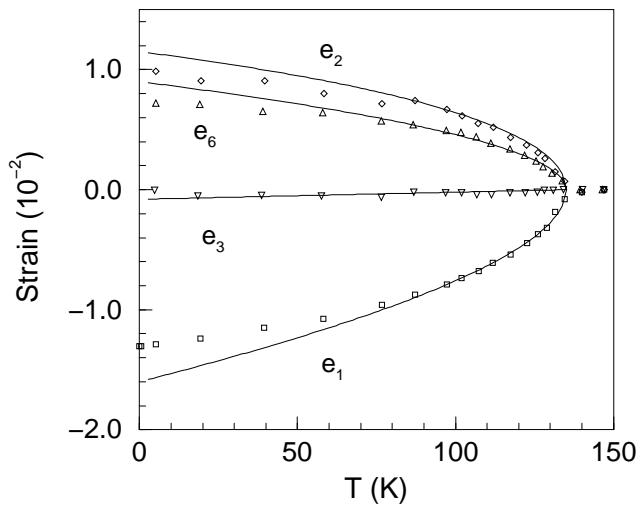}
     \caption{Temperature dependence of the spontaneous strains.
     The points correspond to results obtained from Mr\'oz et al.\cite{Mroz97}
     while the continuous line are calculated using Eq.~\ref{e1me2} -\ref{e6}.
     \label{fig:strains}}
 \end{figure}

So far, our qualitative observations seem to indicate that the
elastic properties of Rb$_4$LiH$_3$(SO$_4$)$_4$ are more consistent
with the predictions based on pseudo-ferroelastic model. A more
rigorous test can be performed by carrying on a detailed numerical
comparison using the prediction giving in the Appendix. While the
elastic constants are determined by sound velocity measurements
realized at room temperature, the magnitude of the coupling terms,
as well as $\alpha$ and $A_4$, need to be determined independently.
Moreover, it is also important to determine the anharmonic
contributions which are independent of the transition. In the case
of Rb$_4$LiH$_3$(SO$_4$)$_4$, these contributions have been
determined from velocity measurements performed on the isomorphic
compound $K_4LiH_3(SO_4)_4$. Finally, let mention that both models
include only eight adjustable parameters and that six of them are
determined independently using the temperature dependence of the
strains, the pressure dependence of $T_c$ ($dT_c/dP$ = 19.1~K/kbar),
and the normalization of the order parameter.  The last two
parameters $\zeta$ and $\eta$, which only influence $C_{44}$ and
$C_{45}$, can be set manually. To illustrate how stringent this
process is, we present in Fig.~\ref{fig:strains} the predictions
based on Eqs.~\ref{e1me2} -\ref{e6} and the temperature dependence
of the strains obtained from high-resolution neutron diffraction
results published by Mr\'oz et al.\cite{Mroz97}
As $e_1 - e_2$ and $e_6$ show a well defined mean field temperature dependence, we can
safely assume that the neutron data reflet the microscopic properties within the
macroscopic domains in the ferroelastic phase. When all adjustable parameters are
determined, we then calculate the velocities along different directions by solving the
Christoffel's equation using the elastic tensors Eq.~\ref{CiiT} and Eq.~\ref{CiiTproper}
for the pseudo-proper and proper ferroelastic transition, respectively.  The results of
the calculations (continuous line) based on the pseudo-proper model are presented in
Fig.~\ref{vvst}-\ref{vvsp} along with the experimental data points. In the para-elastic
phase, the agreement between the prediction of the pseudo-proper model and experimental
results, as a function of temperature and pressure, is quite remarkable (within of
percent). Naturally, the agreement in the ferroelastic phase is not as good considering
that the elastic properties are significantly affected by the existence of structural
domains below T$_c$.  As an ultimate test, we compare in Fig.~\ref{Vsofttheo} the
predictions of the proper model (dash line) and the pseudo-proper model (continuous line)
relative to experimental data showing the largest variation.  These variations are
obtained using transverse modes propagating in the xy plane at an angle $\phi$ relative
to the [110] direction (with $\phi = +5^o,~0^o,~-10^o$) with its polarization in the xy
plane. From that comparison, it is clear that we systematically obtain a better agreement
with the pseudo-proper model rather than the proper ferroelastic model. In
Fig.~\ref{Vsofttheo} we have limited our comparison to $V_{T[\overline{1}10]}[110]$,
however, we reach the same conclusion using results obtained for $V_L[100]$.

\section{Soft mode}
\begin{figure}[tb]
 \includegraphics{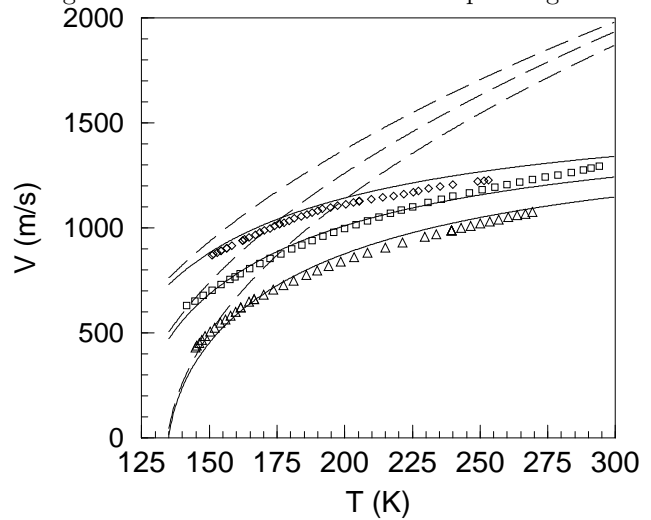}
     \caption{Temperature dependence of the velocity of transverse
     waves polarized in the xy plane and propagating along different directions
     in the xy plane ($\phi = +5^o$, $0^o$, and $-10^o$ relative to the [110] direction).
     Here, the continuous and dash lines correspond to predictions based on a pseudo
     and proper ferroelastic models presented in the Appendix.\label{Vsofttheo} }
\end{figure}
The data presented in Fig.~\ref{Vsofttheo} also confirms a larger
softening as the direction of propagation, for transverse mode
polarized in the xy plane, is a few degrees away from the [110]
direction  (approximately $-10^o$ relative to the $[110]$
direction).  This observation complies with previous Brillouin
scattering measurements\cite{Mroz91} which show less softening as
the direction of propagation is changed toward [010] instead of
[100]. However, in their data analysis, they also claim that
Rb$_4$LiH$_3$(SO$_4$)$_4$ shows incomplete softening at $T_c$, a
claim that we revisit in this section. The effective modulus for the
soft acoustic mode of a $4 \rightarrow 2$ ferroelastic phase
transition can be obtained by finding the eigenvalues of the
submatrix corresponding to the strain components $e_1 - e_2$ and
$e_6$,
\begin{equation}
\left(
  \begin{array}{cc}
  \frac{C_{11} - C_{12}}{2} & C_{16} \\
  C_{16} & C_{66} \\
\end{array}
\right)~~.
 \end{equation}
We would like to point out that the submatrix used here differs from
the one published by Boccara\cite{Boccara68}. Considering that
elastic tensors must be symmetric, it is obvious that some of the
results found in Boccara\cite{Boccara68} are inexact and that they
correspond to simple topographic errors. Thus, contrary to the
expression used by Mroz et al.\cite{Mroz91}, which was derived from
expressions found in Boccara\cite{Boccara68}, the appropriate
relation for the soft modulus is
\begin{widetext}
\begin{equation}\label{eq:softmode}
C_{soft} = \frac{1}{2}\left(\frac{C_{11}-C_{12}}{2}+C_{66} -
\sqrt{\left(\frac{C_{11}-C_{12}}{2}-C_{66}\right)^2+4
C_{16}^2}\right)
\end{equation}
\end{widetext}
while the associated direction of propagation\cite{Dieulesaint} is
given by
\begin{equation}\label{eq:softdir}
    \tan 4 \phi = \frac{4 C_{16}}{C_{11}-C_{12} +2C_{66}}
\end{equation}
where the angle $\phi$ is defined relative the [110] direction.
Using the measured elastic constants of Rb$_4$LiH$_3$(SO$_4$)$_4$ at
room temperature, the expected direction for the soft mode should
correspond to $\phi \cong -13^o$. This prediction ($45^o+\phi =
32^o$ relative to the x-axis) agrees well with the orientations of
the domain walls observed in the ferroelastic phase\cite{Mroz97}. As
that direction does not coincide with any of the principal
crystallographic directions, this has contributed to delay the
experimental determination of the actual soft mode in
Rb$_4$LiH$_3$(SO$_4$)$_4$.  According to our recent investigation,
the data obtained at $\phi = -10^0$ indicate a larger variation
relative to the results obtained at $\phi = 0^0$, which is very
consistent with the numerical prediction. Unfortunately, due to a
rapid increases in the acoustic attenuation close to $T_c$, we were
not able to perform the measurement down to $T_c$. Nevertheless,
considering that the theoretical prediction based on the pseudo
ferroelastic model agree well with our experimental data, we believe
that within the experimental uncertainties there is no strong
evidence of incomplete softening in Rb$_4$LiH$_3$(SO$_4$)$_4$. Thus,
it is clear, based on the results presented in this section, that
the conclusion reached by Mroz et al.\cite{Mroz91} is inexact.

%



\section{Conclusions}

In this investigation, we have presented a detailed analysis of the
elastic properties of Rb$_4$LiH$_3$(SO$_4$)$_4$.  A substantial set
of experimental data, obtained as a function of temperature and
pressure, is compared to two Landau models associated with a $4
\rightarrow 2$ ferroelastic phase transition.  The possible
scenarios considered here correspond to the proper ferroelastic case
where the instability is driven by the strains while in the
pseudo-proper situation the order parameter is unknown.  In the
later case, even the order parameter is unknown, its symmetry must
be identical to the order parameter defined for the proper
ferroelastic case. Both models are derived using a minimum number of
coupling parameters which are adjusted from thermal expansion
data\cite{Mroz97} and the pressure dependence of the critical
temperature $d T_c / d P = 19.1 \pm 0.2~K/kbar$ obtained in this
work. Our analysis clearly shows that within the frame work of the
pseudo-proper ferroelastic model, it is possible to obtain
predictions that are simultaneously consistent with the temperature
and pressure dependence of the elastic constants, the pressure
dependence of $T_c$ and thermal expansion. No such quantitative
agreement is obtained using the proper ferroelastic model.  Thus, as
suggested by our investigation, Rb$_4$LiH$_3$(SO$_4$)$_4$ should be
consider as a pseudo-proper ferroelastic compound.  If our
conclusion is accurate, this implies that the nature of the order
parameter, driving the ferroelastic transition in
Rb$_4$LiH$_3$(SO$_4$)$_4$, is still unknown. So far, there has been
a limited number of Raman scattering measurements on
Rb$_4$LiH$_3$(SO$_4$)$_4$\cite{Mroz92}. However, even if they
observed a soft optical B mode, the authors of Ref.~15 claim that
the variation of that particular mode is insufficient to account for
the transition. It is worthwhile to note that, in the presence of
linear coupling between acoustical and optical
modes\cite{Pinczuk77}, while the actual transition takes place at
T$_c$, the extrapolated temperature dependence of the soft optical
energy in the paraelastic phase is expected to vanish at a
temperature T$_o $. Based on our numerical predictions, $T_o$ is
approximately equal to 20~K in the case of
Rb$_4$LiH$_3$(SO$_4$)$_4$. We believe that further investigations
are certainly necessary in order to clarify the nature of the actual
mode responsible for the ferroelastic transition in
Rb$_4$LiH$_3$(SO$_4$)$_4$.

Finally, as we have clearly demonstrated in this paper, there is no
strong experimental evidence that Rb$_4$LiH$_3$(SO$_4$)$_4$ show
incomplete softening of the soft elastic constant at $T_c$.
Consequently, the ferroelastic character of
Rb$_4$LiH$_3$(SO$_4$)$_4$ is not necessary unique, for example, its
properties seem very similar to those of
(NH$_4$)$_4$LiH$_3$(SO$_4$)$_4$ which shows a $4 \rightarrow 2$
ferroelastic phase transition at 236~K\cite{Mroz93}. It would be
interesting to use a similar numerical analysis in order to
determine the nature of that ferroelastic transition.

\section{Appendix}\label{theory}

\subsection{Landau Models}

The direct group-subgroup relationship for the $4 \rightarrow 2$
ferroelastic phase transition imposes that the spontaneous
strain\cite{Toledano83} associated with this transition must
correspond to the combination $\alpha_1~ (e_1 -e_2) + \alpha_6~
e_6$. This effective strain can then act as the principal order
parameter (proper ferroelastic transition) or couple bilinearly to
the order parameter Q.  In the later case, the order parameter Q
could correspond to the softening of an optical phonon and the phase
transformation is then referred to as a pseudoproper ferroelastic
transition\cite{Toledano83}. Even if there is no direct evidence of
a soft optical mode in Rb$_4$LiH$_3$(SO$_4$)$_4$\cite{Mroz92}, this
later scenario should not be dismissed. As we show in this paper,
these two types of ferroelastic transition should in principle
produce subtle differences in the elastic properties of
Rb$_4$LiH$_3$(SO$_4$)$_4$.  The predictions, derived from the models
presented in the next section, are then compared to our experimental
data in order to determine whether Rb$_4$LiH$_3$(SO$_4$)$_4$ should
be regarded as a proper or pseudo-proper ferroelastic compound.

\subsection{Pseudo-proper Ferroelastic Model} \label{pseudo}

Considering that we are particularly interested in the pressure and
temperature dependence of the elastic properties, we express the
Gibbs free energy into four distinct contributions such as
\begin{equation}\label{eq:Ftotal}
G(Q, e_i) = F_L(Q) + F_{el}(e_i) + F_c(Q, e_i) + F_P(P, e_i)~.
\end{equation}
Here, $F_L(Q)$ is the usual second-order Landau-type free energy in
terms of an order parameter Q,
\begin{eqnarray}\label{eq:Lq}
    F_L(Q)  &=& \frac{1}{2}~A_2~Q^2 +\frac{1}{4}~ A_4~Q^4
\end{eqnarray}
where $A_2 = \alpha~ (T-T_o)$ depends explicitly on temperature while the pressure
contribution is introduced in $F_P(P, e_i)$ of Eq.~\ref{eq:Ftotal}. In the presence of
external stresses $\sigma_i$, this contribution is normally written as $ - \sum_i
\sigma_i~ e_i$. In the case of an hydrostatic pressure P the stress associated with
longitudinal strains ($i$ = 1, 2, 3) corresponds to $\sigma_i = - P$ while for shear
strains $\sigma_j = 0$ ($j$ = 4, 5, 6), hence
\begin{equation}\label{eq:Fp}
F_P(P, e_i) = P~(e_1 + e_2 + e_3).
\end{equation}
This contribution is crucial in order to calculate the pressure
dependence of the elastic constants as well as to determine the
pressure dependence of the critical temperature.  Furthermore, this
is achieved without adding any additional adjustable parameters. The
elastic energy $F_{el}(e_i)$ is also straightforward to derive as it
is imposed by the symmetry of the high temperature phase. Thus, for
a point group 4 symmetry\cite{Dieulesaint}
\begin{eqnarray}\label{eq:Fel}
    F_{el}(e_i)&=& \frac{1}{2}~C_{11}~(e_1^2+e_2^2)+\frac{1}{2}~C_{33}~e_3^2
    +\frac{1}{2}~C_{44}~(e_4^2+e_5^2) \nonumber \\
    &+&\frac{1}{2}~C_{66}~e_6^2+C_{12}~e_1~e_2+C_{13}~(e_1+e_2)~e_3\nonumber \\
    &+&C_{16}~(e_1-e_2)~e_6
\end{eqnarray}
where C$_{ik}$ represent the bare elastic constants at high
temperatures.  Finally, the most critical term corresponds to the
coupling energy $F_c(Q, e_i)$ which takes into account the coupling
between the strain components $e_i$ and the order parameter Q. For a
$4 \rightarrow 2$ pseudo-proper ferroelastic transition, we know
that the order parameter $Q$ must transform as the spontaneous
strain $\alpha_1 (e_1 -e_2) + \alpha_6 e_6$ under the symmetry
operations of the high temperature phase.  Thus, using fundamental
symmetry arguments, the coupling terms considered in this paper are
\begin{widetext}
\begin{eqnarray}\label{eq:Fc}
    F_c(Q, e_i) & = &  \beta~Q~(e_1-e_2)+ \gamma~Q~e_6 +\delta~Q^2~e_3
     +  \lambda~Q^2~(e_1 + e_2)+ \zeta~ e_4~ e_5~ Q + \eta~ (e_4^2 + e_5^2)~ Q^2
\end{eqnarray}
\end{widetext}
Here, we only consider terms which are fully compatible with a $4 \rightarrow 2$ structural
transformation. The Gibbs energy, as defined in this paper (Eq.~\ref{eq:Ftotal}-\ref{eq:Fc}), gives a phenomenological
framework which can now be used in order to derive the elastic properties of a $4 \rightarrow 2$ pseudo-proper
ferroelastic compound. From the minimization of the Gibbs energy with respect to $e_i$, we obtain a series of
expressions for the spontaneous strains $e_i(Q)$. At ambient pressure these relations correspond to
\begin{eqnarray}
&e_1-e_2& = \frac{2~(\gamma~ C_{16} - \beta~ C_{66})}{(C_{11}-C_{12})~C_{66}-2~C_{16}^2} ~Q \label{e1me2} \\
&e_1+e_2& = \frac{2~  (\delta~ C_{13} - \lambda~ C_{33})}{(C_{11}+C_{12})~C_{33}-2~C_{13}^2}~Q^2\label{e1pe2} \\
&e_3& = -\frac{  (\delta~ (C_{11}+ C_{12})- 2~\lambda~ C_{13})}{(C_{11}+C_{12})~C_{33}-2~C_{13}^2} ~Q^2 \label{e3}\\
&e_4& = 0 \\
&e_5& = 0 \\
&e_6& = -\frac{\gamma~(C_{11}-C_{12})-2~\beta~
C_{16}}{(C_{11}-C_{12})~C_{66}-2~C_{16}^2} ~Q \label{e6}
\end{eqnarray}
As $(e_1 - e_2)$ and $e_6$ display the same symmetry characteristic
as Q, it is then natural to find that they are proportional to Q. We
also obtain that $e_4 = e_5 = 0$ which is again compatible with a $
4 \rightarrow 2$ symmetry change. Furthermore, using the
minimization with respect to the order parameter Q, the
corresponding expressions for the order parameter Q, the critical
temperature $T_c$ and its pressure dependence can be written as
\begin{eqnarray}\label{dTcdP}
Q(T,P) & = & \sqrt{\frac{\alpha~ C_a~(T_c + \frac{d T_c}{d P}P - T )}{\Delta}}\\
T_c & = & T_o-\frac{C_b}{\alpha~C_a} \\
  \frac{d T_c}{d P} & = & 2 \frac{C_c}{\alpha~C_a}
\end{eqnarray}
where
\begin{widetext}
\begin{eqnarray}
C_a & = & (C_{11}+C_{12})~C_{33}-2~ C_{13}^2\\
C_b & = &\gamma^2~(C_{11}-C_{12})+2\beta~(\beta~C_{66} -2\gamma~C_{16})\\
C_c & = &\delta~(C_{11}+C_{12}-2~C_{13})+2\lambda~(C_{33} - C_{13})\\
\Delta & = & 8 \delta \lambda C_{13} - 2 A_4 C_{13}^2 - 4 \lambda^2
C_{33} + C_{11}(A_4 C_{33} - 2 \delta^2) + C_{12} (A_4 C_{33}-2
\delta^2)
\end{eqnarray}
Finally, the elastic constants are calculated using\cite{Rehwald73}
\begin{eqnarray}
C_{mn}^*& =& \frac{\partial^2 F}{\partial e_m \partial e_n}
  -  \frac{\partial^2 F}{\partial Q \partial e_m}\left(\frac{\partial^2
F}{\partial Q^2 }\right)^{-1}\frac{\partial^2 F}{\partial e_n
\partial Q}~.
\end{eqnarray}
According to this model, the elastic tensor associated with the
tetragonal 4 phase is
\begin{equation}\label{CiiT}
\left(
    \begin{array}{cccccc}
    C_{11}-\frac{\beta^2}{A(T,P)} & C_{12}+\frac{\beta^2}{A(T,P)} & C_{13} & 0 & 0 & C_{16}-\frac{\beta \gamma}{A(T,P)} \\
    C_{12}+\frac{\beta^2}{A(T,P)} & C_{11}-\frac{\beta^2}{A(T,P)} & C_{13} & 0 & 0 & -C_{16}+\frac{\beta \gamma}{A(T,P)} \\
    C_{13} & C_{13} & C_{33} & 0 & 0 & 0 \\
    0 & 0 & 0 & C_{44} & 0 & 0 \\
    0 & 0 & 0 & 0 & C_{44} & 0 \\
    C_{16}-\frac{\beta \gamma}{A(T,P)} & - C_{16}+\frac{\beta \gamma}{A(T,P)} & 0 & 0 & 0 & C_{66}-\frac{\gamma^2}{A(T,P)} \\
\end{array}
\right)
\end{equation}
while the corresponding tensor for the monoclinic 2 phase is
\begin{equation}\label{CiiTlow}
\left(
    \begin{array}{cccccc}
    C_{11}-\frac{X_{Q_+}^2}{Y_Q} & C_{12}+\frac{X_{Q_+}X_{Q_-}}{Y_Q}& C_{13}-\frac{2 \lambda Q X_{Q_+}}{Y_Q} & 0 & 0 & C_{16}-\frac{\gamma X_{Q_+}}{Y_Q} \\
    C_{12}+\frac{X_{Q_+}X_{Q_-}}{Y_Q} & C_{11}-\frac{X_{Q_-}^2}{Y_Q}  & C_{13}+\frac{2 \lambda Q X_{Q_-}}{Y_Q} & 0 & 0 & -C_{16}+\frac{\gamma X_{Q_-}}{Y_Q} \\
    C_{13}-\frac{2 \lambda Q X_{Q_+}}{Y_Q} & C_{13}+\frac{2 \lambda Q X_{Q_-}}{Y_Q} & C_{33}-\frac{4 \delta^2 Q^2}{Y_Q} & 0 & 0 & -\frac{2\gamma \delta Q}{Y_Q} \\
    0 & 0 & 0 & C_{44}+2 \eta Q^2 & \zeta Q & 0 \\
    0 & 0 & 0 & \zeta Q  & C_{44}+2 \eta Q^2 & 0 \\
    C_{16}-\frac{\gamma X_{Q_+}}{Y_Q} &-C_{16}+\frac{\gamma X_{Q_-}}{Y_Q} & -\frac{2\gamma \delta Q}{Y_Q} & 0 & 0 & C_{66}-\frac{\gamma^2}{Y_Q} \\
\end{array}
\right)
\end{equation}
with
\begin{eqnarray}
A(T, P) & = & \alpha \left(T - T_o - \frac{d T_c}{d P}P\right)\\
X_{Q_+} & = & \beta + 2 \lambda Q \\
X_{Q_-} & = & \beta - 2 \lambda Q \\
Y_Q &=&2A_4Q^2-\frac{\gamma^2 (C_{11}-C_{12})- 2 \beta^2
C_{66}}{(C_{11}-C_{12})C_{66} - 2 C_{16}^2}~.
\end{eqnarray}
\end{widetext}
A rapid inspection of the elastic tensor, corresponding to the solution below $T_c$ (or above $P_c$), indicates that
its form is consistent with the elastic tensor of the monoclinic class  2. This indicates that the coupling terms
considered in the Gibbs energy includes all terms which are fully compatible with a $4 \rightarrow 2$ symmetry change.

\subsection{Proper Ferroelastic Model} \label{proper}

In the case of a proper ferroelastic model, the derivation is slightly different.  We can still write the Gibbs energy,
the Landau energy, and the elastic energy as defined in Eq.~\ref{eq:Ftotal} - Eq.~\ref{eq:Lq}, and Eq.~\ref{eq:Fel},
respectively.  The main difference being that the order parameter Q now corresponds to a strain combination defined as
$Q = \alpha_1 (e_1 - e_2) + \alpha_6 e_6 $. Thus, the coupling terms invariant with respect to the high temperature
symmetry are
\begin{widetext}
\begin{eqnarray}\label{eq:Fcproper}
    F_c(e_i) & = &  \delta~Q^2~e_3 + \lambda~Q^2~(e_1 + e_2)+ \zeta~ e_4~ e_5~ Q
    + \eta~ (e_4^2 + e_5^2)~ Q^2
\end{eqnarray}
Using the minimization of the Gibbs energy with respect to $e_i$, the solutions for $e_1+ e_2$, $e_3$, $e_4$, $e_5$,
$dT_c/dP$ remain identical to those obtained for the pseudo-proper case while
\begin{equation}\label{Tcproper}
    T_c=T_o-\frac{-2C_{16}^2+(C_{11}-C_{12})C_{66}}{\alpha(\alpha_6^2 (C_{11}-C_{12})+
    2 \alpha_1 (\alpha_1 C_{66} - 2 \alpha_6 C_{16})}~~.
\end{equation}
Finally, as that the Gibbs energy is now only a function of strains,
the elastic constants are directly obtained using
\begin{eqnarray}
C_{mn}^*& =& \frac{\partial^2 F}{\partial e_m ~\partial e_n}~~.
\end{eqnarray}
Thus, in the case a $4 \rightarrow 2$ proper ferroelastic transition, the elastic tensor for the paraelastic phase is
given by
\begin{equation}\label{CiiTproper}
\left(
    \begin{array}{cccccc}
    C_{11}+\alpha_1^2 A(T,P) & C_{12}-\alpha_1^2 A(T,P) & C_{13} & 0 & 0 & C_{16}+\alpha_1 \alpha_6 A(T,P) \\
    C_{12}-\alpha_1^2A(T,P) & C_{11}+\alpha_1^2 A(T,P) & C_{13} & 0 & 0 & -C_{16}-\alpha_1 \alpha_6 A(T,P) \\
    C_{13} & C_{13} & C_{33} & 0 & 0 & 0 \\
    0 & 0 & 0 & C_{44} & 0 & 0 \\
    0 & 0 & 0 & 0 & C_{44} & 0 \\
    C_{16}+ \alpha_1 \alpha_6 A(T,P) & - C_{16}- \alpha_1 \alpha_6 A(T,P) & 0 & 0 & 0 & C_{66}+\alpha_6^2A(T,P) \\
\end{array}
\right)
\end{equation}
\end{widetext}
where $A(T, P) = \alpha \left(T - T_c - \frac{d T_c}{d P}P\right)$. Thus, the temperature and pressure dependence of
the elastic constants should be linear.  This differs from the pseudo-proper model where the variation in the elastic
constants is expected to be inversely proportional to $A(T,P)$. Naturally, depending on the strength of the coupling
parameters this dependence could be quasi-linear.  Thus, a clearly distinction between these two models can only be
obtained through a detailed comparison with experimental results.

\section{Acknowledgments}
This work was supported by grants from the Natural Science and Engineering Research
Council of Canada (NSERC), Canada Foundation for Innovation (CFI) as well as  by the
Ministry of Education and Science (Poland), Grant No. 1 PO3B 066 30.


%





\end{document}